\documentclass[aps,prl,twocolumn,groupedaddress,showpacs]{revtex4-1}
\usepackage{graphicx} 
\usepackage{color} 
\newcommand{\mut}{\tilde{\mu}}
\newcommand{\zetat}{\tilde{\zeta}}
\newcommand{\lambdat}{\tilde{\lambda}}

\begin{document}

\title{Colloids electrophoresis for strong and weak ion diffusivity}

\author{Giovanni Giupponi}
\email{giupponi@ffn.ub.es}
\author{Ignacio Pagonabarraga}
\email{ipagonabarraga@ub.edu}
\affiliation{Departament de Fisica Fonamental, Universitat de Barcelona, Carrer Mart\'i i Franques, 08028 Barcelona (Spain)}
\date{\today}

\begin{abstract}
We study the electrophoretic flow of suspensions of charged
colloids with a  mesoscopic method that allows to model generic
experimental conditions. We show that for highly charged colloids  their
electrophoretic mobility increases significantly  and displays a mobility
maximum on increasing the colloidal charge  for all salt
concentrations. The electrophoretic mobility of highly charged colloids is also
enhanced significantly when ion advection is dominant, leading to  a strong
heterogeneity in the local electrophoretic response especially at low salt
concentration, when ion diffuse layers overlap.
\end{abstract}


\maketitle

Effective electrostatic interactions between charged colloidal particles in
solution play a fundamental role determining the macroscopic phase and
rheological properties of colloidal suspensions that are pivotal for many
applications in material science~\cite{Russel91, Larson99}. In particular, the
response of these systems to applied electric fields enhances the degree of
experimental control and tunability of suspensions down to the nanoscale and provides
a natural means to design and operate nanodevices. As a result, electrokinetic
techniques are pragmatically exploited to control fluid flow at small scales, as for
example in nano-fluidic molecular sorting~\cite{Sparreboom09, Austin07},
nano-medicine protocols~\cite{Andre04} or micro-pattern assembly\cite{Hayward00}.

A theoretical and quantitative knowledge of electrophoresis is
still not complete as understanding the electrophoretic flow (EF) of the solvent
and the charged components (ionized macromolecules, counter- and salt ions)
requires the combined treatment of solvent flow coupled to the diffusion and
advection of the dissolved electrolyte in the presence of charged colloids.
The dynamical coupling between suspended charged components  through
hydrodynamics, leading to effective electrostatic interactions,  does not allow
for  exact solutions; analytic predictions are restricted to the linear regime
and either numerical or simulation studies are required to assess the interplay
between electrostatics and hydrodynamics that results in the EF.  The relevant
(length$|$time) scales involved in EF range from the
nano-(meter$|$second), characteristic of the charge distribution around
the particles, to the colloidal size itself, up to the micro-(meter$|$second).
This disparity of scales complicates the computational modeling because
molecular dynamics is practically limited to the detailed resolution of the
smallest, nano-scales. Mesoscopic modeling (MM)~\cite{Hoogerbrugge92, Benzi92,
Malevanets99}, which includes the appropriate solvent dynamics at a coarse
scale, overcomes the difficulties of handling multi-scale simulations of a
charged colloidal suspension EF. MM has been recently used  to study the
non-equilibrium dynamics of a variety of complex charged he\-te\-ro\-ge\-neous
systems~\cite{Pagonabarraga10},  and therefore offers a valuable way to obtain
theoretical predictions for a comprehensive EF, \emph{in primis} the
fundamental EF of a charged spherical particle.

In this Letter, we study the electrokinetic response of charged colloids and
examine the relevance of non-linear coupling and ion specificity for different
types of EF due to charge density deformations induced by applied
electric fields. We do so by taking advantage of a novel mesoscopic
simulation technique that naturally includes appropriate boundary conditions
and does not require any assumption beyond Poisson-Boltzmann. 
We then  capture the significant physical couplings between charges,
colloids and solvent, accounting for the non-linear response of the
electrolyte rearrangements to the applied field. Moreover, our model can be
used under general experimental conditions of salt concentration, ion
diffusivity and colloid surface charge density.

A charged macromolecule  in solution is surrounded by
a counter- and salt ion cloud  which screens the colloidal charge depending
on the salt concentration, $c_s$, and the solvent  dielectric properties. Applying an
electric field $E$ causes the electrolyte and macromolecule to  move. The limiting particle 
velocity at the steady state, $v_{l}$, develops as a result of the balance between the
electrostatic and viscous forces. However, the deformed charge
distribution of co- and counter-ions around the  macromolecule, which gives
rise to an electric double layer (EDL) around the  object,  is hard to derive
analytically even in thermodynamic equilibrium and usually solutions are known
for  low electrostatic potentials at the colloid surface (referred to as the  zeta potential,
    $\zeta$), when linear electrostatics holds~\cite{Russel91, BCSCL03}.
Moreover, non-linear EDL distortion due to the external electric field  leads
to further difficulties in its analytic understanding.

O' Brien and White (OW)~\cite{OBW78} have
integrated numerically the electrokinetic equations for the EF of an infinitely diluted
spherical colloid in the linear regime. Their results indicate that the
particle electrophoretic mobility, $\mu=v_{l}/E$, depends  on the colloid
radius $a$, salt concentration $c_s$ and dielectric properties of the solvent
at temperature $T$ through the Debye screening length, $\lambda_D = (8\pi l_b
    c_s z_+^2)^{-1/2}$ (for a symmetric electrolyte of valence $z_+$, here $z_+=
      \pm1$, in a solvent with Bjerrun length $l_b=e^2/(4\pi\epsilon
        k_BT)$~\cite{BCSCL03}, with  $e$  the electron charge and  $k_B$  the
      Boltzmann factor) and $\zeta$, predicting  either
    mobility saturation or a maximum for high $\zeta$ values. 
Using different mesoscopic models, Kim {\sl et al.}~\cite{KimPRL06} were able to
confirm OW predictions~\cite{OBW78} for low $c_s$, while
Lobaskin {\sl et al.}~\cite{HolmPRL07etal, Dunweg08etal} showed that the
electrophoretic behaviour of salt-free systems can be systematically mapped to
a corresponding low-salt suspension. Both cases, however, address the restricted
case of small $\zeta$ and $c_s$, where OW predictions   are expected to hold.

Our mesoscopic representation of charged  colloidal suspensions builds on a
continuous description of the electrolyte, characterized  in terms of the anion
and cation local densities, $\rho_{\pm}$. These densities evolve according to
the electrokinetic  equations, which  read \begin{equation}
\label{eq1}
\frac{\partial \rho_k}{\partial t} = - \nabla \cdot \rho_k\vec{v} + \nabla \cdot D_k[\nabla \rho_k + e\beta z_k \rho_k \nabla \varphi],
\end{equation}
\begin{equation}
\label{eq2} 
\frac{\partial \rho \vec{v}}{\partial t} = \eta \nabla^2 \vec{v} - \nabla p_{id}+ \beta\sum_k e z_k\rho_k \nabla \varphi,
\end{equation}
\begin{equation}
\label{eq3} 
\nabla^2 \varphi = - \frac{1}{\epsilon}\sum_k e z_k\rho_k
\end{equation}
where $D_k, z_k, k=+,-$ are the diffusivities and
valences of positive and negative ions, $\rho, \vec{v}, p_{id}$ and $ \eta$ correspond to  the
solvent density, velocity, ideal pressure and shear viscosity, $\varphi$ is the electrostatic
potential and  $\epsilon$ the solvent permittivity. 
Eq.~\ref{eq1} expresses ion  mass  conservation as a result of  diffusion and
advection, while the solvent motion, Eq.~\ref{eq2}, evolves according   to the Navier-Stokes equation for a viscous fluid accelerated by     electrostatic forces due to local charge density. Finally,  the Poisson
equation enforces the electrostatic coupling between the charged species and
the macromolecules. 

The solvent motion  emerges from  a discrete lattice
formulation of Boltzmann's kinetic equation~\cite{Benzi92} coupled to a
discrete solution of the convection-diffusion equation for the dissolved charged ion 
species~\cite{Capuani04}. Hence, we regard the
counter- and salt ions as scalar fields within Poisson-Boltzmann (PB)
level~\cite{BCSCL03}.  The colloidal macromolecules are individually resolved,
  embedded on the lattice and coupled to the fluid through appropriate kinetic
  rules applied on their boundaries~\cite{Ladd06I}.  The hydrodynamic forces
  exerted by the fluid on the suspended particles, together with the
  electrostatic and dispersion forces
determine the motion of the macromolecules in the fluid~\cite{Rotenberg10}.
The  finite resolution of the colloidal particles on a lattice requires a
proper calibration to identify the effective size where $\zeta$ is consistent
with the colloidal charge~\cite{dsfd_rome}. The electrostatic potential drop
around the colloids emerges consistently as a result of the ionic dynamics
coupled to the fluid flow without further assumptions.  Although such an
approach disregards ion correlations, it provides a general
framework to address electrokinetics  at weak and strong couplings and
identifies the relevant competing physical mechanisms in charged driven fluids. 

\begin{figure}
\includegraphics[width=7.8cm]{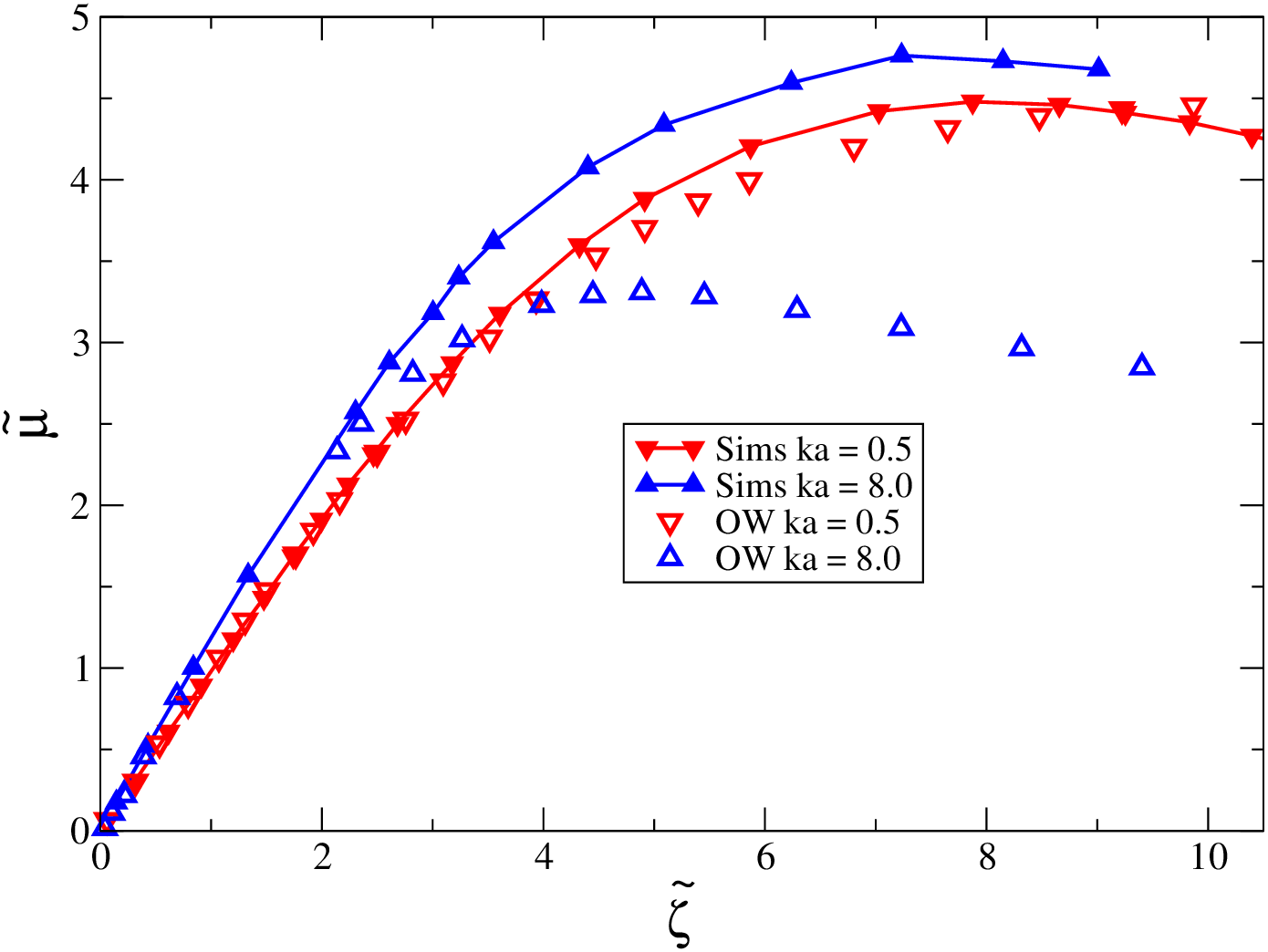}
\caption{Dimensionless electrophoretic mobility for a spherical colloidal particle, $\mut$, 
versus the dimensionless zeta potential,$\zetat$. Results from simulations (filled
    symbols) and OW\cite{OBW78} (empty symbols) for $ka=0.5(8.0)$, triangles down(up).\label{fig1}}
\end{figure}

We start by analyzing the electrophoretic mobility, $\mu$, of a crystal of
spherical colloids of charge $Ze$ and radius $a$ as a function of $c_s$, $\zeta$ and volume
fraction, $\Phi = 4\pi a^3/3L^3$, where $L$ stands for the system size. In
order to compare with experimental and numerical results, we introduce the
dimensionless mobility, $\mut = \frac{6\pi\mu\eta l_B}{e}$, and  zeta potential,
$\zetat = \frac{e\zeta}{k_BT}$. The units in  the simulations  ensure that the
relevant scales keep the  right ordering and are properly resolved on the
lattice~\cite{lb_unitsetal}. We take $\kappa a = 0.5, 8.0$, with
$\kappa=1/\lambda_D$,  representative of low to high salt
concentrations~\cite{OBW78}, covering  two orders of
magnitude in molarity, $\sim 10^{-4}-10^{-2}$ M, when mapping our simulation systems to
polystyrene spheres with a radius of $17$ nm~\cite{Palberg04aetal, *Palberg04b,
*Palberg04cetal}. The applied electric field is tuned to remain
in the linear response regime, $E \ll \zetat /
\lambda_D$.

Fig.~\ref{fig1} shows the dimensionless mobility, $\mut$, against  $\zetat$.
For small values, $\zetat \lesssim 3$, we observe an excellent agreement with
OW theoretical prediction~\cite{OBW78}, which holds in
the linear regime for low values of $\zetat$. Our data show that increasing
$\zetat$ leads to an enhancement of $\mut$ with respect to OW. Such a deviation
increases with $\zetat$ and with the amount of charge in the system up to
$50\%$ for narrow double layers,  $\kappa a=8.0$. Although the theoretical
predictions from OW display a maximum  outside the linear regime and only for
thin double layers ($\kappa a \gtrsim 2.75$), the maxima observed in
Fig.~\ref{fig1}  appear at higher zeta potentials and develop for any $\kappa a$.  
Therefore, these results indicate that care has to be taken when using OW
curves to compare with  experimental results.

\begin{figure}
\includegraphics[width=7.8cm]{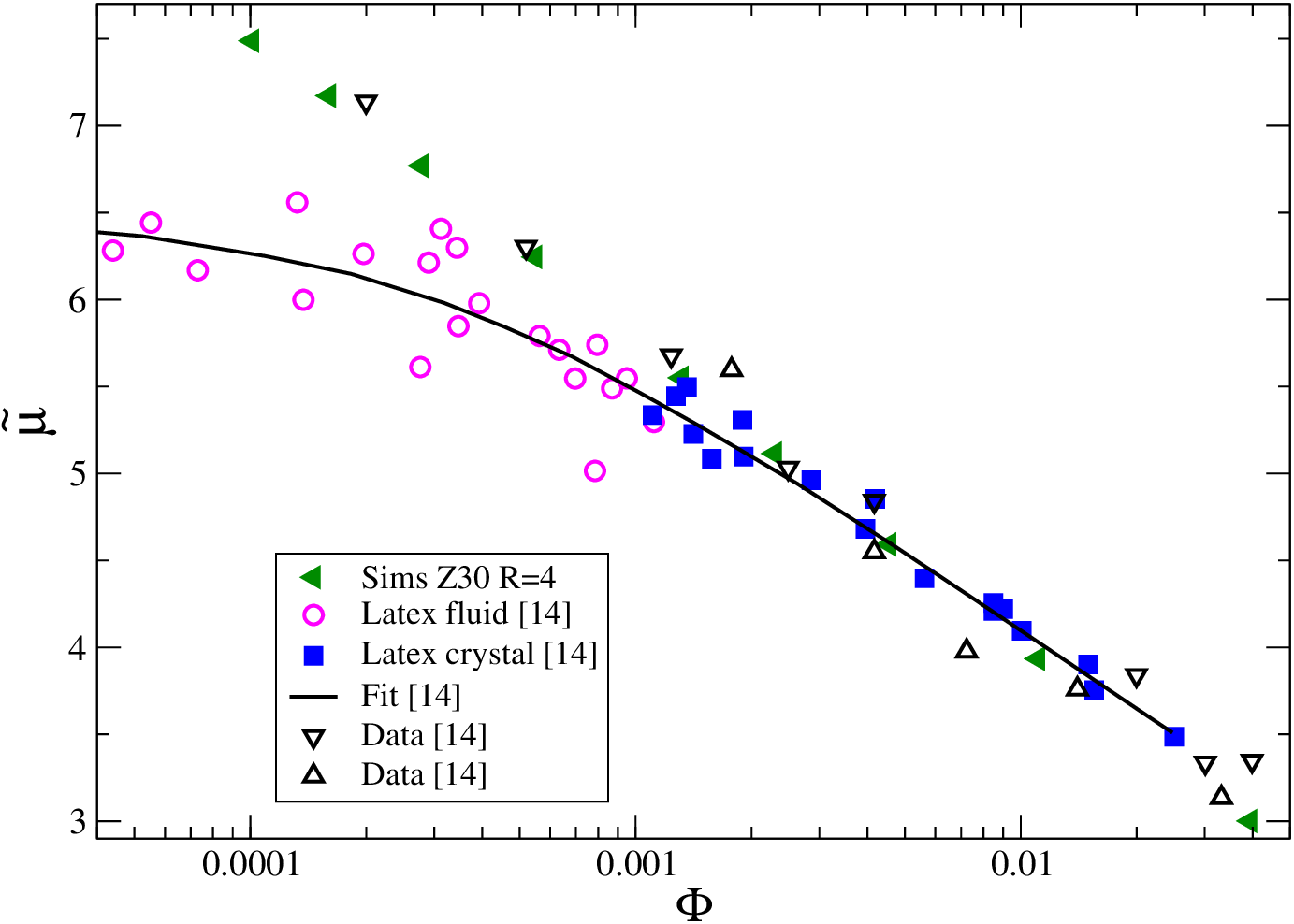}
\caption{Dimensionless electrophoretic mobility $\mut$ for a colloidal
particle with charge $Z=30$ and radius $a=4.0$ versus colloid volume fraction
$\Phi$ (triangle left). No salt is added to the system. Experimental (sphere and square symbols) and
simulation results (empty triangles) from Lobaskin {\sl et
al.}\cite{HolmPRL07etal}\label{fig2}.}
\end{figure}

The model described reproduces quantitatively   experimental data for low
$\zetat$, salt-free colloidal suspensions. In
Fig.~\ref{fig2}, we show that agreement with experimental data is obtained for
latex particles suspended in water\cite{HolmPRL07etal, Palberg04aetal, *Palberg04b,
*Palberg04cetal} for colloid volume fractions $\Phi \gtrsim 0.001$.
For very low $\Phi$, the ionic contribution due to water dissociation  is
relevant~\cite{HolmPRL07etal}  and the mobility dependence on  $\Phi$  vanishes.
Fig.~\ref{fig2} also displays  simulation results  where counterions are
resolved individually~\cite{HolmPRL07etal}. The agreement shows  that  the
electrophoretic response observed experimentally can be  captured through the
counterion density and that  ionic finite-size  electrostatic correlations
are subdominant.

\begin{figure}
\includegraphics[width=7.8cm]{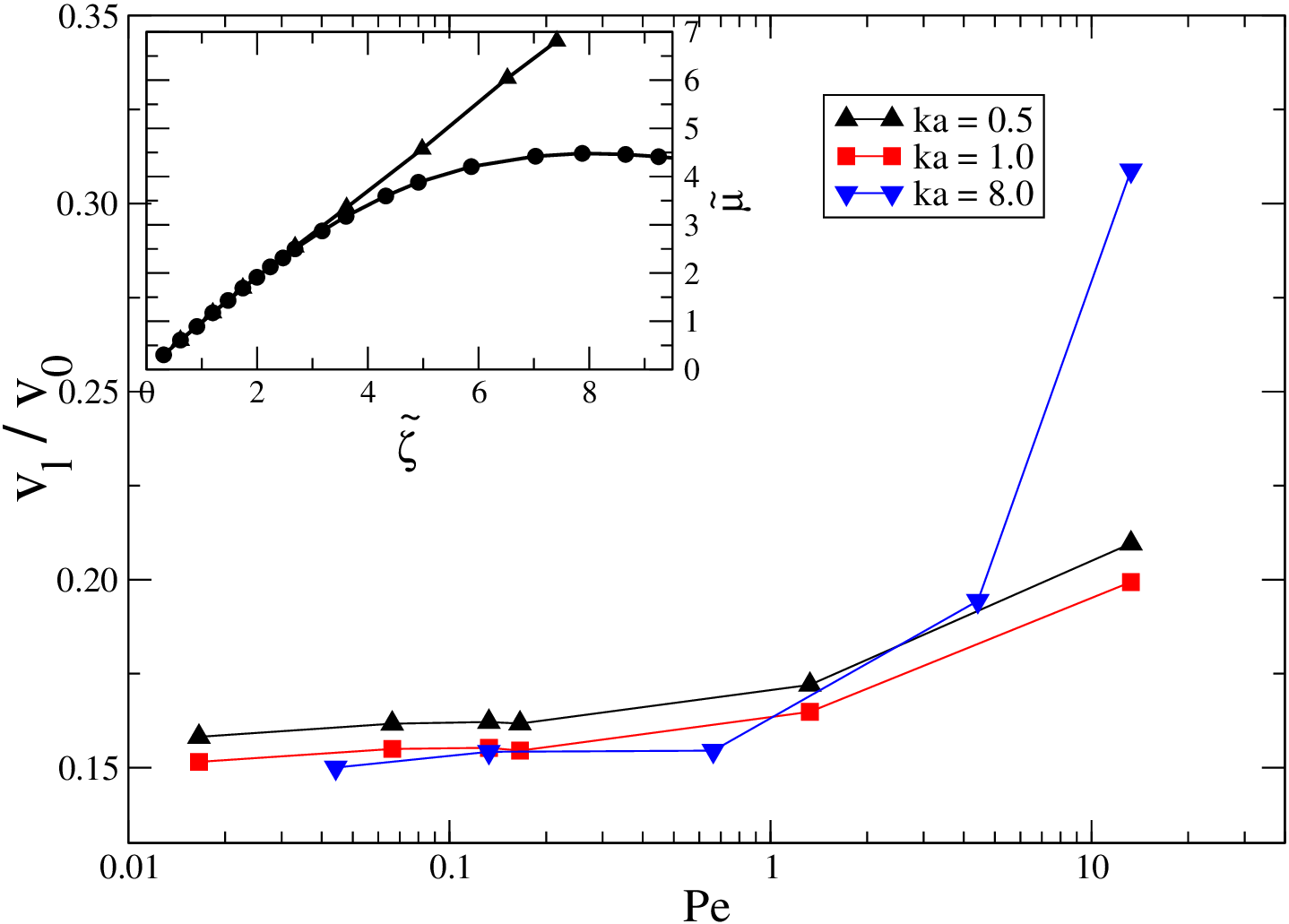}
\caption{Steady state colloidal velocity $v_{l}/v_0$ versus P\'eclet flow number
for $\kappa a=0.5, 1.0, 8.0$.Inset: dimensionless mobility $\mut$
versus dimensionless zeta-potential, $\zetat$, for $\kappa a=0.5$, $Pe$ $\sim
10^{-3}$ (filled circles) and $Pe \sim 0.3$ (filled
triangles)\label{fig3}.}
\end{figure}

Even if in many situations individual ion resolution is irrelevant, its diffusion may affect
significantly the electrokinetic response of charged colloids.  We have
analyzed   the effect of ion diffusion for  symmetric monovalent
electrolytes, $D=D_{\pm}$, and have quantified the importance of electrolyte
diffusion to flow advection  in terms of the  P\'eclet number, $Pe=av_0/D$,
where the characteristic fluid velocity is induced by the colloid and
reads  $v_0 =  e Z E /6\pi  \eta a$.  Fig.~\ref{fig3} shows the
dimensionless steady state velocity $v_{l}/v_0$ for a colloidal crystal at $\Phi
\sim 10^{-2}$ as a function of $Pe$ for different salt concentrations $c_s$,
$\kappa a=0.5, 1.0, 8.0$, covering from narrow to  wide EDL at a fixed
colloidal charge density, $\zetat \sim 5.0$, in the non-linear regime. The
results show that for strong ion diffusivity the mobility is not affected by
advection, while at larger $Pe$ a  significant increase in the colloidal
mobility, departing from the  high diffusivity  regime at $Pe=0$,  is observed.
The inset of Fig.~\ref{fig3}, where we display $\mut$ as a function of the
colloidal $\zetat$ for regimes where ion diffusivity is either subdominant or
relevant, indicates that the competition between ion diffusion and advection by
the incoming flow becomes quantitatively significant in the non-linear regime.
The mobility enhancement reported at  small
electric fields arises from  ion  advection, therefore
differing from the mobility enhancement associates to charge
stripping under strong applied  fields\cite{Fixman83}. Although we cannot
discard that a maximum in $\mut$ develops asymptotically, for the range of
$\zetat$ accessible experimentally, $\zetat \lesssim 10$\cite{Russel91,OBW78},
  we do not observe a saturation of $\mut$. Results in
  Figs.~\ref{fig1}-\ref{fig2} correspond to the small $Pe \sim 10^{-3}$ regime
  to compare with theories and experiments where diffusivity is dominant;
  however the results of Fig.~\ref{fig3} indicate that diffusion can become  a
  relevant variable because a wide range of $Pe$ can be 
 reached experimentally. For example, for a colloidal particle of radius
 $a\sim 0.1$ $\mu m$,  charge density $\sigma \sim 8\cdot 10^{-2} C/m^2$ and an
 external field  $E_{exp} \sim 10^3
V/m$~\cite{Palberg04aetal, *Palberg04b, *Palberg04cetal}, $Pe\sim
10^{-1}$~\cite{Russel91} for an electrolyte composed of small ions with typical
diffusivity, $D \sim 10^{-9} m^2/s$,
 $Pe$ can nevertheless increase one or two orders of magnitude imposing strong
electric fields or in the presence of bigger, nano-sized ions~\cite{Faraudo09etal}
which induce a  greater EDL disruption. To characterize the EDL degree of
distortion  as a function of $Pe$, we  compute the dimensionless eigenvalues
$\lambdat$ of the charge density inertia tensor around the colloid center,
$\lambdat_i =\lambda_i/(Za^2)$, $i=x,y,z$. If we apply the external electric
field along the $x$ axis, the axisymmetric structure of the EDL gives an
asymmetry of the transverse eigenvalue as $Pe$ increases. 
\begin{table}[b]
\caption{Charge  intertia matrix eigenvalues.}
\label{Ltable}
\begin{ruledtabular}
\begin{tabular}{cccccc}
$\kappa a$ & $Pe$& $\lambdat_x$ & $\lambdat_y$ & $\lambdat_z$ & $\mut$\\
\hline
0.5 & $2.54\cdot10^{-3}$ & 1.626848 & 1.627185 & 1.627185 & 4.42 \\
0.5 & $3.12\cdot10^{-1}$ & 1.676013 & 1.676420 & 1.676420 & 6.81 \\
\hline
8.0 & $6.63\cdot10^{-3}$ & 0.756485 & 0.756834 & 0.756834 & 4.34 \\
8.0 & $8.60\cdot10^{-1}$ & 0.764143 & 0.764346 & 0.764346 & 5.62 \\
\end{tabular}
\end{ruledtabular}
\end{table}
Table~\ref{Ltable} reports the
magnitude of the three eigenvalues for different values of $Pe$  and $c_s$.  For all salt
concentrations $\lambdat_i$ increase with $Pe$, showing a larger spread of the
EDL that results in a lower screening of the colloid charge, therefore
raising its mobility at steady state.  The different values of  $\lambdat_i$
show that the axisymmetric charge distribution around the colloid develops a
significant departure from isotropy only when ion  advection drag becomes
dominant, at  experimentally high $Pe$.

\begin{figure}
\includegraphics[width=7.8cm]{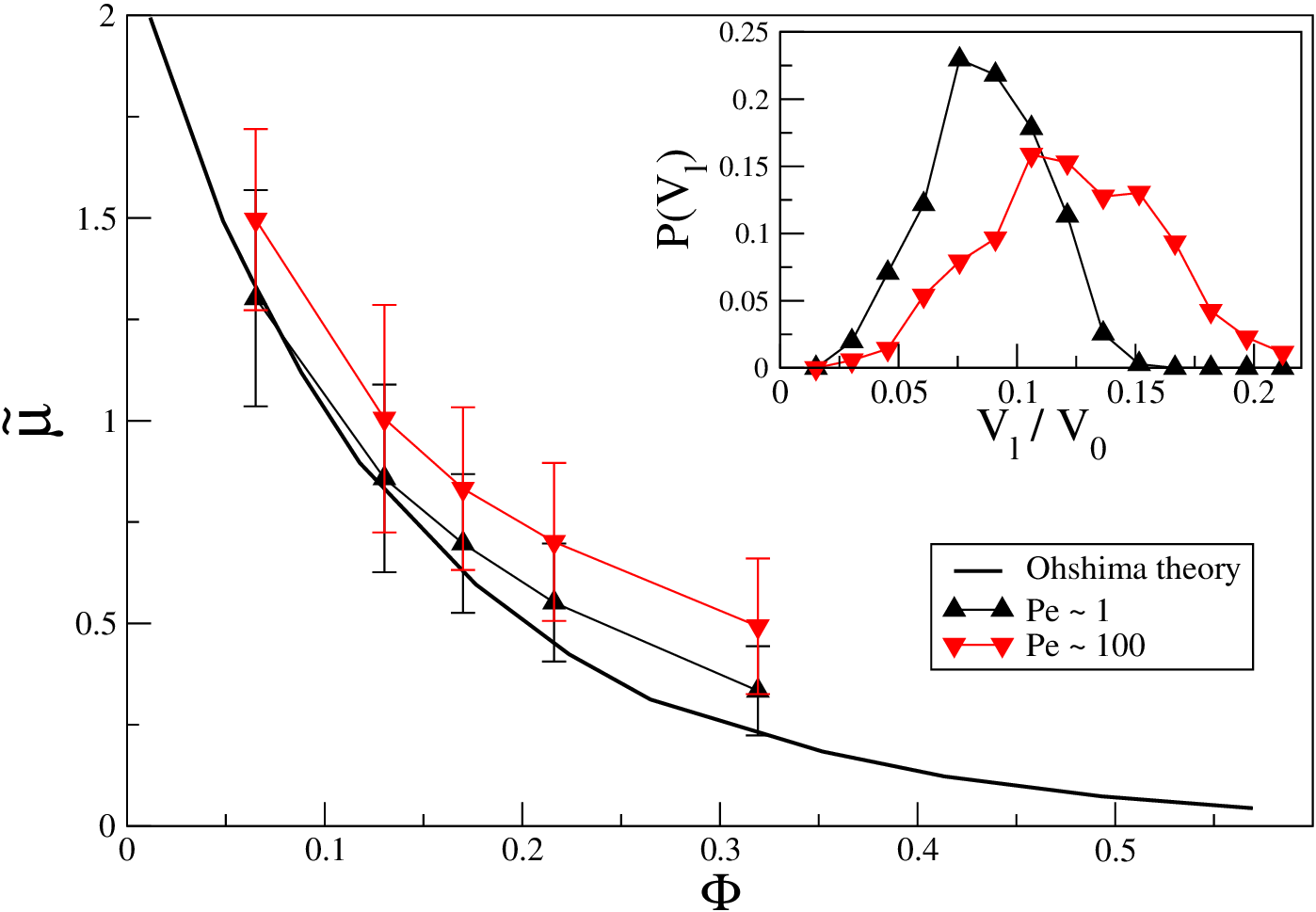}
\caption{Dimensionless mobility $\mut$ versus volume fraction $\Phi$ for a
suspension of particles with overlapping EDLs, $Pe\sim1, 10^2$.  Inset:
probability distribution of particle velocities $v_l$ when $\Phi=0.32,
\kappa a=1.0$ (overlapping EDLs).
\label{fig4}}
\end{figure}

Finally, we study colloidal suspensions at low salt concentrations and at  
volume fractions when EDLs overlap. We compare our results for $\zetat=2.0$
with analytical predictions derived by Ohshima using single-particle
charge distributions and linear electrostatics~\cite{Ohshima06}. 
Fig.~\ref{fig4} shows that deviations from Ohshima's theory develop for $\Phi \simeq
0.1$, in agreement with the theoretical expectation that EDL overlap 
for $\Phi \geq \kappa a/(1+\kappa a)^3$.  We unveil also 
the fundamental role played by advection-diffusion competition in EF, as $Pe$
leads to an increase in the electrophoretic mobility. Moreover, ion diffusivity
is seen to alter conjointly  the dispersion of $v_l$  in the linear regime for
significant  EDL overlap, as depicted in the inset of Fig.~\ref{fig4}.
Understanding the physical mechanisms underlying the  
increase in  $\tilde{\mu}$ with $\Phi$ for  highly diluted suspensions with overlapping EDLs~\cite{Palberg04aetal},
$\Phi< 10^{-3}$, requires further
investigation beyond current computational capabilities. The probability
distribution function (pdf) of colloid velocities for $\Phi=0.32$, $\kappa a
=1.0$ shows that on increasing $Pe$ not only the mean $v_l$ shifts to larger
values but also the pdf widens.  As a result, an increase in ion drag due to EF
leads to larger spread in local $v_l$, enhancing the colloidal local
electrophoretic response heterogeneity. 

In summary, we have comprehensively studied the electrokinetic response of
suspensions of charged colloids, opening up new possibilities to
fine tune and extend current electrokinetic techniques. We have shown that the
colloidal mobility develops a maximum due to the competition between EDL
distortion and the EF induced by the applied electric field. We have revealed
the importance of ion diffusivity identifying an experimental regime of
enhanced mobility and increased
heterogeneity of local electrophoretic response when the external driving
dominates over ion diffusion. At small ion diffusivities, the overlap of EDL
leads to a strong dynamic coupling between colloids, causing strong
heterogeneities in the colloidal response.
We acknowledge J. Faraudo for fruitful discussions, the IEF Marie Curie
scheme (GG) and the Direcci\'on General de Investigaci\'on (Spain) and DURSI
project (IP) for financial support under projects  FIS\ 2008-04386 and
2009SGR-634, respectively.

\bibliography{elb}

\begin{thebibliography}{10}%
\makeatletter
\providecommand \@ifxundefined [1]{%
 \ifx #1\undefined \expandafter \@firstoftwo
 \else \expandafter \@secondoftwo
\fi
}%
\providecommand \@ifnum [1]{%
 \ifnum #1\expandafter \@firstoftwo
 \else \expandafter \@secondoftwo
\fi
}%
\providecommand \enquote [1]{``#1''}%
\providecommand \bibnamefont  [1]{#1}%
\providecommand \bibfnamefont [1]{#1}%
\providecommand \citenamefont [1]{#1}%
\providecommand\href[0]{\@sanitize\@href}%
\providecommand\@href[1]{\endgroup\@@startlink{#1}\endgroup\@@href}%
\providecommand\@@href[1]{#1\@@endlink}%
\providecommand \@sanitize [0]{\begingroup\catcode`\&12\catcode`\#12\relax}%
\@ifxundefined \pdfoutput {\@firstoftwo}{%
 \@ifnum{\z@=\pdfoutput}{\@firstoftwo}{\@secondoftwo}%
}{%
 \providecommand\@@startlink[1]{\leavevmode\special{html:<a href="#1">}}%
 \providecommand\@@endlink[0]{\special{html:</a>}}%
}{%
 \providecommand\@@startlink[1]{%
  \leavevmode
  \pdfstartlink
   attr{/Border[0 0 1 ]/H/I/C[0 1 1]}%
   user{/Subtype/Link/A<</Type/Action/S/URI/URI(#1)>>}%
  \relax
 }%
 \providecommand\@@endlink[0]{\pdfendlink}%
}%
\providecommand \url  [0]{\begingroup\@sanitize \@url }%
\providecommand \@url [1]{\endgroup\@href {#1}{\urlprefix}}%
\providecommand \urlprefix [0]{URL }%
\providecommand \Eprint[0]{\href }%
\@ifxundefined \urlstyle {%
  \providecommand \doi [1]{doi:\discretionary{}{}{}#1}%
}{%
  \providecommand \doi [0]{doi:\discretionary{}{}{}\begingroup
  \urlstyle{rm}\Url }%
}%
\providecommand \doibase [0]{http://dx.doi.org/}%
\providecommand \Doi[1]{\href{\doibase#1}}%
\providecommand \bibAnnote [3]{%
  \BibitemShut{#1}%
  \begin{quotation}\noindent
    \textsc{Key:}\ #2\\\textsc{Annotation:}\ #3%
  \end{quotation}%
}%
\providecommand \bibAnnoteFile [2]{%
  \IfFileExists{#2}{\bibAnnote {#1} {#2} {\input{#2}}}{}%
}%
\providecommand \typeout [0]{\immediate \write \m@ne }%
\providecommand \selectlanguage [0]{\@gobble}%
\providecommand \bibinfo [0]{\@secondoftwo}%
\providecommand \bibfield [0]{\@secondoftwo}%
\providecommand \translation [1]{[#1]}%
\providecommand \BibitemOpen[0]{}%
\providecommand \bibitemStop [0]{}%
\providecommand \bibitemNoStop [0]{.\EOS\space}%
\providecommand \EOS [0]{\spacefactor3000\relax}%
\providecommand \BibitemShut [1]{\csname bibitem#1\endcsname}%
\bibitem{Russel91}%
  \BibitemOpen
  \bibfield{author}{%
  \bibinfo {author} {\bibfnamefont{W.~B.}\ \bibnamefont{Russel}}, \bibinfo
  {author} {\bibfnamefont{D.~A.}\ \bibnamefont{Saville}},\ and\ \bibinfo
  {author} {\bibfnamefont{W.~R.}\ \bibnamefont{Schowalter}},\ }%
  \emph{\bibinfo {title} {{Colloidal dispersions}}}\ (\bibinfo {publisher}
  {Cambridge University Press},\ \bibinfo {year} {1991})%
  \bibAnnoteFile{NoStop}{Russel91}%
\bibitem{Larson99}%
  \BibitemOpen
  \bibfield{author}{%
  \bibinfo {author} {\bibfnamefont{R.~G.}\ \bibnamefont{Larson}},\ }%
  \emph{\bibinfo {title} {{The structure and rheology of complex fluids}}}\
  (\bibinfo {publisher} {Oxford University Press},\ \bibinfo {year} {1999})%
  \bibAnnoteFile{NoStop}{Larson99}%
\bibitem{Sparreboom09}%
  \BibitemOpen
  \bibfield{author}{%
  \bibinfo {author} {\bibfnamefont{W.}~\bibnamefont{Sparreboom}}, \bibinfo
  {author} {\bibfnamefont{A.}~\bibnamefont{Van Den~Berg}},\ and\ \bibinfo
  {author} {\bibfnamefont{J.~C.~T.}\ \bibnamefont{Eijkel}},\ }%
  \bibfield{journal}{%
  \bibinfo {journal} {Nature Nanotech.}\ }%
  \textbf{\bibinfo {volume} {4}},\ \bibinfo {pages} {713} (\bibinfo {year}
  {2009})%
  \bibAnnoteFile{NoStop}{Sparreboom09}%
\bibitem{Austin07}%
  \BibitemOpen
  \bibfield{author}{%
  \bibinfo {author} {\bibfnamefont{R.}~\bibnamefont{Austin}},\ }%
  \bibfield{journal}{%
  \bibinfo {journal} {Nature Nanotech.}\ }%
  \textbf{\bibinfo {volume} {2}},\ \bibinfo {pages} {79} (\bibinfo {year}
  {2007})%
  \bibAnnoteFile{NoStop}{Austin07}%
\bibitem{Andre04}%
  \BibitemOpen
  \bibfield{author}{%
  \bibinfo {author} {\bibfnamefont{F.}~\bibnamefont{Andre}}\ and\ \bibinfo
  {author} {\bibfnamefont{L.~M.}\ \bibnamefont{Mir}},\ }%
  \bibfield{journal}{%
  \bibinfo {journal} {Gene therapy}\ }%
  \textbf{\bibinfo {volume} {11}},\ \bibinfo {pages} {S33} (\bibinfo {year}
  {2004})%
  \bibAnnoteFile{NoStop}{Andre04}%
\bibitem{Hayward00}%
  \BibitemOpen
  \bibfield{author}{%
  \bibinfo {author} {\bibfnamefont{R.~C.}\ \bibnamefont{Hayward}}, \bibinfo
  {author} {\bibfnamefont{D.~A.}\ \bibnamefont{Saville}},\ and\ \bibinfo
  {author} {\bibfnamefont{I.~A.}\ \bibnamefont{Aksay}},\ }%
  \bibfield{journal}{%
  \bibinfo {journal} {Nature}\ }%
  \textbf{\bibinfo {volume} {404}},\ \bibinfo {pages} {56} (\bibinfo {year}
  {2000})%
  \bibAnnoteFile{NoStop}{Hayward00}%
\bibitem{Hoogerbrugge92}%
  \BibitemOpen
  \bibfield{author}{%
  \bibinfo {author} {\bibfnamefont{P.~J.}\ \bibnamefont{Hoogerbrugge}}\ and\
  \bibinfo {author} {\bibfnamefont{J.}~\bibnamefont{Koelman}},\ }%
  \bibfield{journal}{%
  \bibinfo {journal} {Europhys. Lett.}\ }%
  \textbf{\bibinfo {volume} {19}},\ \bibinfo {pages} {155} (\bibinfo {year}
  {1992})%
  \bibAnnoteFile{NoStop}{Hoogerbrugge92}%
\bibitem{Benzi92}%
  \BibitemOpen
  \bibfield{author}{%
  \bibinfo {author} {\bibfnamefont{R.}~\bibnamefont{Benzi}}, \bibinfo {author}
  {\bibfnamefont{S.}~\bibnamefont{Succi}},\ and\ \bibinfo {author}
  {\bibfnamefont{M.}~\bibnamefont{Vergassola}},\ }%
  \bibfield{journal}{%
  \bibinfo {journal} {Phys. Rep.}\ }%
  \textbf{\bibinfo {volume} {222}},\ \bibinfo {pages} {145} (\bibinfo {year}
  {1992})%
  \bibAnnoteFile{NoStop}{Benzi92}%
\bibitem{Malevanets99}%
  \BibitemOpen
  \bibfield{author}{%
  \bibinfo {author} {\bibfnamefont{A.}~\bibnamefont{Malevanets}}\ and\ \bibinfo
  {author} {\bibfnamefont{R.}~\bibnamefont{Kapral}},\ }%
  \bibfield{journal}{%
  \bibinfo {journal} {J. Chem. Phys.}\ }%
  \textbf{\bibinfo {volume} {110}},\ \bibinfo {pages} {8605} (\bibinfo {year}
  {1999})%
  \bibAnnoteFile{NoStop}{Malevanets99}%
\bibitem{Pagonabarraga10}%
  \BibitemOpen
  \bibfield{author}{%
  \bibinfo {author} {\bibfnamefont{I.}~\bibnamefont{Pagonabarraga}}, \bibinfo
  {author} {\bibfnamefont{B.}~\bibnamefont{Rotenberg}},\ and\ \bibinfo {author}
  {\bibfnamefont{D.}~\bibnamefont{Frenkel}},\ }%
  \bibfield{journal}{%
  \bibinfo {journal} {Phys. Chem. Chem. Phys.}\ }%
  \textbf{\bibinfo {volume} {12}},\ \bibinfo {pages} {9566} (\bibinfo {year}
  {2010})%
  \bibAnnoteFile{NoStop}{Pagonabarraga10}%
\bibitem{BCSCL03}%
  \BibitemOpen
  \bibfield{author}{%
  \bibinfo {author} {\bibfnamefont{J.}~\bibnamefont{Barrat}}\ and\ \bibinfo
  {author} {\bibfnamefont{J.}~\bibnamefont{Hansen}},\ }%
  \emph{\bibinfo {title} {{Basic concepts for simple and complex liquids}}}\
  (\bibinfo {publisher} {Cambridge University Press},\ \bibinfo {year} {2003})%
  \bibAnnoteFile{NoStop}{BCSCL03}%
\bibitem{OBW78}%
  \BibitemOpen
  \bibfield{author}{%
  \bibinfo {author} {\bibfnamefont{R.~W.}\ \bibnamefont{O'Brien}}\ and\
  \bibinfo {author} {\bibfnamefont{L.~R.}\ \bibnamefont{White}},\ }%
  \bibfield{journal}{%
  \bibinfo {journal} {J. Chem. Soc., Faraday Transactions 2}\ }%
  \textbf{\bibinfo {volume} {74}},\ \bibinfo {pages} {1607} (\bibinfo {year}
  {1978})%
  \bibAnnoteFile{NoStop}{OBW78}%
\bibitem{KimPRL06}%
  \BibitemOpen
  \bibfield{author}{%
  \bibinfo {author} {\bibfnamefont{K.}~\bibnamefont{Kim}}, \bibinfo {author}
  {\bibfnamefont{Y.}~\bibnamefont{Nakayama}},\ and\ \bibinfo {author}
  {\bibfnamefont{R.}~\bibnamefont{Yamamoto}},\ }%
  \bibfield{journal}{%
  \bibinfo {journal} {Phys. Rev. Lett.}\ }%
  \textbf{\bibinfo {volume} {96}},\ \bibinfo {pages} {208302} (\bibinfo {year}
  {2006})%
  \bibAnnoteFile{NoStop}{KimPRL06}%
\bibitem{HolmPRL07etal}%
  \BibitemOpen
  \bibfield{author}{%
  \bibinfo {author} {\bibfnamefont{V.}~\bibnamefont{Lobaskin}} \emph{et~al.},\
  }%
  \bibfield{journal}{%
  \bibinfo {journal} {Phys. Rev. Lett.}\ }%
  \textbf{\bibinfo {volume} {98}},\ \bibinfo {pages} {176105} (\bibinfo {year}
  {2007})%
  \bibAnnoteFile{NoStop}{HolmPRL07etal}%
\bibitem{Dunweg08etal}%
  \BibitemOpen
  \bibfield{author}{%
  \bibinfo {author} {\bibfnamefont{B.}~\bibnamefont{D\"unweg}} \emph{et~al.},\
  }%
  \bibfield{journal}{%
  \bibinfo {journal} {J. Phys.: Cond. Matt.}\ }%
  \textbf{\bibinfo {volume} {20}},\ \bibinfo {pages} {404214} (\bibinfo {year}
  {2008})%
  \bibAnnoteFile{NoStop}{Dunweg08etal}%
\bibitem{Capuani04}%
  \BibitemOpen
  \bibfield{author}{%
  \bibinfo {author} {\bibfnamefont{F.}~\bibnamefont{Capuani}}, \bibinfo
  {author} {\bibfnamefont{I.}~\bibnamefont{Pagonabarraga}},\ and\ \bibinfo
  {author} {\bibfnamefont{D.}~\bibnamefont{Frenkel}},\ }%
  \bibfield{journal}{%
  \bibinfo {journal} {J. Chem. Phys.}\ }%
  \textbf{\bibinfo {volume} {121}},\ \bibinfo {pages} {973} (\bibinfo {year}
  {2004})%
  \bibAnnoteFile{NoStop}{Capuani04}%
\bibitem{Ladd06I}%
  \BibitemOpen
  \bibfield{author}{%
  \bibinfo {author} {\bibfnamefont{A.~J.~C.}\ \bibnamefont{Ladd}},\ }%
  \bibfield{journal}{%
  \bibinfo {journal} {J. Fluid Mech.}\ }%
  \textbf{\bibinfo {volume} {271}},\ \bibinfo {pages} {285} (\bibinfo {year}
  {2006})%
  \bibAnnoteFile{NoStop}{Ladd06I}%
\bibitem{Rotenberg10}%
  \BibitemOpen
  \bibfield{author}{%
  \bibinfo {author} {\bibfnamefont{B.}~\bibnamefont{Rotenberg}}, \bibinfo
  {author} {\bibfnamefont{I.}~\bibnamefont{Pagonabarraga}},\ and\ \bibinfo
  {author} {\bibfnamefont{D.}~\bibnamefont{Frenkel}},\ }%
  \bibfield{journal}{%
  \bibinfo {journal} {Faraday Disc.}\ }%
  \textbf{\bibinfo {volume} {144}},\ \bibinfo {pages} {223} (\bibinfo {year}
  {2010})%
  \bibAnnoteFile{NoStop}{Rotenberg10}%
\bibitem{dsfd_rome}%
  \BibitemOpen
  \bibfield{author}{%
  \bibinfo {author} {\bibfnamefont{G.}~\bibnamefont{Giupponi}}\ and\ \bibinfo
  {author} {\bibfnamefont{I.}~\bibnamefont{Pagonabarraga}},\ }%
  \bibfield{journal}{%
  \bibinfo {journal} {Phil. Trans. R. Soc. A}\ }%
  \textbf{\bibinfo {volume} {369}},\ \bibinfo {pages} {2546} (\bibinfo {year}
  {2011})%
  \bibAnnoteFile{NoStop}{dsfd_rome}%
\bibitem{lb_unitsetal}%
  \BibitemOpen
  \bibfield{author}{%
  \bibinfo {author} {\bibfnamefont{M.}~\bibnamefont{Cates}} \emph{et~al.},\ }%
  \bibfield{journal}{%
  \bibinfo {journal} {J. Phys: Condens. Matter}\ }%
  \textbf{\bibinfo {volume} {16}} (\bibinfo {year} {2004})%
  \bibAnnoteFile{NoStop}{lb_unitsetal}%
\bibitem{Palberg04aetal}%
  \BibitemOpen
  \bibfield{author}{%
  \bibinfo {author} {\bibfnamefont{N.}~\bibnamefont{Garbow}} \emph{et~al.},\ }%
  \bibfield{journal}{%
  \bibinfo {journal} {J. Phys.: Cond. Matt.}\ }%
  \textbf{\bibinfo {volume} {16}},\ \bibinfo {pages} {3835} (\bibinfo {year}
  {2004})%
  \bibAnnoteFile{NoStop}{Palberg04aetal}%
\bibitem{Palberg04b}%
  \BibitemOpen
  \bibfield{author}{%
  \bibinfo {author} {\bibfnamefont{M.}~\bibnamefont{Medebach}}\ and\ \bibinfo
  {author} {\bibfnamefont{T.}~\bibnamefont{Palberg}},\ }%
  \bibfield{journal}{%
  \bibinfo {journal} {J. Phys.: Cond. Matt.}\ }%
  \textbf{\bibinfo {volume} {16}},\ \bibinfo {pages} {5653} (\bibinfo {year}
  {2004})%
  \bibAnnoteFile{NoStop}{Palberg04b}%
\bibitem{Palberg04cetal}%
  \BibitemOpen
  \bibfield{author}{%
  \bibinfo {author} {\bibfnamefont{T.}~\bibnamefont{Palberg}} \emph{et~al.},\
  }%
  \bibfield{journal}{%
  \bibinfo {journal} {J. Phys.: Cond. Matt.}\ }%
  \textbf{\bibinfo {volume} {16}},\ \bibinfo {pages} {4039} (\bibinfo {year}
  {2004})%
  \bibAnnoteFile{NoStop}{Palberg04cetal}%
\bibitem{Fixman83}%
  \BibitemOpen
  \bibfield{author}{%
  \bibinfo {author} {\bibfnamefont{M.}~\bibnamefont{Fixman}}\ and\ \bibinfo
  {author} {\bibfnamefont{S.}~\bibnamefont{Jagannathan}},\ }%
  \bibfield{journal}{%
  \bibinfo {journal} {Macromolecules}\ }%
  \textbf{\bibinfo {volume} {16}},\ \bibinfo {pages} {685} (\bibinfo {year}
  {1983})%
  \bibAnnoteFile{NoStop}{Fixman83}%
\bibitem{Faraudo09etal}%
  \BibitemOpen
  \bibfield{author}{%
  \bibinfo {author} {\bibfnamefont{A.}~\bibnamefont{Mart\'{\i}n-Molina}}
  \emph{et~al.},\ }%
  \bibfield{journal}{%
  \bibinfo {journal} {Soft Matter}\ }%
  \textbf{\bibinfo {volume} {5}},\ \bibinfo {pages} {329} (\bibinfo {year}
  {2009})%
  \bibAnnoteFile{NoStop}{Faraudo09etal}%
\bibitem{Ohshima06}%
  \BibitemOpen
  \bibfield{author}{%
  \bibinfo {author} {\bibfnamefont{H.}~\bibnamefont{Ohshima}},\ }%
  \emph{\bibinfo {title} {{Theory of colloid and interfacial electric
  phenomena}}}\ (\bibinfo {publisher} {Academic Press},\ \bibinfo {year}
  {2006})%
  \bibAnnoteFile{NoStop}{Ohshima06}%
\end{thebibliography}%
\end{document}